\documentstyle[twoside,epsfig,fleqn,espcrc2]{article}

\title{On the Character of Leading Asymmetry in the Hadroproduction of 
Charmed Mesons and Baryons.}

\author{O.I.Piskounova,\\
          P.N.Lebedev Physical Institute, Leninski pr. 53, 117924 Moscow, Russia}

\begin{document}

\begin{abstract}

The character of asymmetry between the spectra of $\Lambda_c$ and ${\bar{\Lambda}}_c$,
obtained recently in the E781 experiment (FNAL) is discussed on the basis 
of the Quark Gluon String Model(QGSM).
As it was shown in the description of the asymmetry between the spectra of 
leading and nonleading charmed mesons measured in $\Sigma^--A$ interactions at $p_L$= 340 GeV/c in the
WA89 experiment in previous studies, the asymmetries between 
$D^-$ and $D^+$ meson spectra and between $D_s^-$ and $D^+_s$ meson spectra 
can be fitted by QGSM curves obtained with the same parameter of string
fragmentation, $a_1$=10, as well as the asymmetry between the D-meson spectra
in $\pi^-A$ collisions at the E791 experiment. The forms of $\Lambda_c$/$\bar{\Lambda_c}$ 
asymmetry dependenses measured in $\Sigma^-A$ and $p-A$ collisions 
at $p_L$= 600 GeV/c in the E781 experiment are different. It is shown in the 
framework of QGSM that they depend on whether the diquarks of beam and 
target particles took part in charmed baryon formation or not. The QGSM 
results are compared with the calculations carried out by the other 
authors.
\end{abstract}
\vspace{1pc}

\maketitle

\section{INTRODUCTION}

The difference in $x$ spectra (x=$x_F=2p_{II}/\sqrt{s}$) of leading and
nonleading particles  has been discussed recently and several theoretical
models explained successfully the asymmetry as an effect of
an interplay between the quark contents of the projectile
and of the produced hadron. The charmed particles of leading type (LP) containing 
at least one ordinary quark of the same type as the beam particle
have higher average $x$ value than the particles (NLP) of the same sort having no
one quark in common with the projectile . 

The asymmetry A(x) is usually defined as:
\begin{equation}
A(x)=\frac{dN^{LP}/dx-dN^{NLP}/dx}{dN^{LP}/dx+dN^{NLP}/dx}.
\end{equation}
Two experiments have measured recently the asymmetries between the 
spectra of $D^-$ and $D^+$, $D_s^-$
and $D^+_s$ mesons as well as between $\Lambda_c$ and $\bar{\Lambda_c}$ 
in $\Sigma^-A$ interactions: WA89  \cite{wa89} (CERN)
at $p_L$= 340 GeV/c and E781 \cite{selex}(FNAL) at $p_L$= 600 GeV/c.
It seems to be interesting to consider these results from the point of view
of Quark Gluon String Model (QGSM) and to compare with $x_F$ asymmetry 
dependencies obtained in the $\pi^-A$ experiments \cite{wa92,e791}in 
order to understand the influence of quark composition of beam 
particle on the features of heavy flavored particle production. 

 The nonperturbative approach accepted in our QGSM model \cite{qgsm} exploits 
the properties of fragmentation functions in order to insert the asymmetry.

\section{QUARK-GLUON STRING MODEL}

\subsection{The Quark Distributions in QGSM}

The process of multiparticle production can be illustrated in QGSM with the cut
n-pomeron diagram.
The inclusive production cross section of hadrons H is written as
a sum over n-Pomeron cylinder diagrams:
\begin{equation}
f_{1}= \int E \frac{d^{3}\sigma^{H}}
{d^{3}p}d^{2}p_{\bot}=\sum_{n=0}^{\infty}\sigma_{n}(s) \varphi_{n}^{H}(s,x)
\end{equation}
where function  $\varphi_{n}^{H}(s,x)$ is a particle distribution in the
configuration of n cut cylinders and $\sigma_{n}$ is the probability of
this process. The parameter of the supercritical Pomeron used here
is $\Delta_P=\alpha_P(0)-1=0.12$. 

The distribution functions are discribed in previous papers \cite{kaons,prev papers}.
It should be mentioned here that a structure function of i-th quark wich has a
fraction of energy $x_{1}$ in the interacting hadron,$f^i(x_1)$, and ${\cal D}_{i}^{H}(z)$
,a fragmentation function of this quark into the considered type
of produced hadrons H, are constracted according to the  Regge 
asymptotic rules proposed in \cite{kaidalov}. 
In the case of hyperon beam they depend
on the parameter of the Regge trajectory of $\varphi$-mesons ($s\bar{s}$)
because of s-quark contained in $\Sigma^-$.

\subsection{The Fragmentation Functions.}

 The following favoured fragmentation function into $D_{s}^{-}$-mesons
was written, for instance, for the strange valence quark fragmentation:
\begin{equation}
{\cal D}_{s}^{D_{s}^{-}}(z)=\frac{1}{z}(1-z)^{-\alpha_{\psi}(0)+\lambda}
(1+a_{1}^{D_{s}}z^{2}),
\end{equation}
where  $\lambda$=2$\alpha_{D^{*}}^{\prime}$(0)$\bar{p_{\bot D^{*}}^{2}}$.
An additional factor $(1+a_{1}^{D_{s}}z^{2})$
provides the parametrization of the probability of heavy
quark production in the interval z=0 to z $\rightarrow$ 1. 
The function for the nonleading fragmentation of d-quark chain into 
$D^{+}$ has an additional $(1-z)^{2(1-\alpha_{R}(0))}$ according to 
the same rules:
\begin{equation}
{\cal D}_{d}^{D^{+}}(z)=\frac{1}{z}(1-z)^{-\alpha_R(0)+\lambda+
2(1-\alpha_{R}(0))+\Delta_{\psi}},
\end{equation}
where $\Delta_{\psi}=\alpha_{R}(0)-\alpha_{\psi}$(0).

The dd-diquark fragmentation includes the constant $a_f$ which could be 
interpreted as "leading" parameter, but this value is fixed due to 
baryon number sum rule and should be approximately equal to $a_f$ taken for 
$\Lambda_c$ spectra in our previous calculations \cite{charm}:
\begin{eqnarray}
{\cal D}_{f dd}^{\Lambda_c}(z)&=&\frac{a_f}{a_0z}z^{2\alpha_R(0)-2\alpha_N(0)} \\
  &&(1-z)^{-\alpha_{\psi}(0)+\lambda+2(1-\alpha_R(0)} \nonumber
\end{eqnarray}
where the term $z^{2\alpha_R(0)-2\alpha_N(0)}$ means the probability for
initial diquark to have z close to 0.

\subsection{The Distributions and the Fragmentation Functions of Sea Quarks.}

Some fractions of sea quark pairs in hyperon, $d\bar{d}$ and $s\bar{s}$, are to 
be taken into account as far as they suppress the leading/nonleading asymmetry. 
The structure functions of ordinary
quark pairs in the quark sea of hyperon can be written by the same way 
as the valence quark distributions.

As soon as we accounted $d\bar{d}$ and $s\bar{s}$ fraction in the quark 
sea of hyperon some fraction of charmed sea quark are to be 
considered as well. This small heavy quark admixture 
plays an important role due to its strong impact on the difference 
between leading and nonleading charmed meson spectra. 

The charmed sea quark structure function is similar to the 
distribution of strange sea quarks \cite{kaons}:
\begin{eqnarray}
 f_{\Sigma^{-}}^{c,\bar{c}}(x_{1})=C_{c,\bar{c}}^{(n)}
   \delta_{c,\bar{c}}x_{1}^{-\alpha_{\psi}(0)} \\
   (1-x_{1})^{\alpha_{R}(0)-2\alpha_{N}(0)+\Delta_{\varphi}
    +\Delta_{\psi}+n-1+2(1-\alpha_{R}(0))} \nonumber
\end{eqnarray}
where $\delta_{c,\bar{c}}$ is the weight of charm admixture in 
the quark sea of hyperon. In fact it is not necessarily to be equal to 
the charmed quark fraction in quark sea of pion \cite{pionbeam}. This is 
only one parameter we
can vary for $\Sigma^{-}$ interaction after the best fit of pion
experimental data which had been done before. The value 
of $\delta_{c,\bar{c}}$ 
can be estimated in the description of the WA89 data on $D_s$- and 
$D$-meson asymmetries.

Fragmentation functions into $D$ mesons are the following:
\begin{equation}
{\cal D}_{c,\bar{c}}^{D^{+},D^{-}}(z)=\frac{1}{z}z^{1-\alpha_{\psi}(0)}(1-z)^
{-\alpha_{R}(0)+\lambda}.
\end{equation}

\section{EXPERIMENTAL DATA ON ASYMMETRY}

\subsection{$D^-$/$D^+$ and $D_s^-$/$D_s^+$ asymmetries.}

The main parameter of QGSM scheme which is responsible for 
leading/nonleading charm asymmetry is $a_1$, defined above. The fraction of charmed sea quarks, $\delta_{(c,\bar{c})}$, 
is the second parameter in this calculations which makes the asymmetry 
lower because of the equal amounts of $D^+$ and $D^-$ mesons produced by 
each sea charmed quark pair. Two sets of this couple of parameters were
chosen in the description of $\pi^-A$ reaction data: $a_1$=4, $\delta_{(c,
\bar{c})}$=0 and  $a_1$= 10, $\delta_{(c,\bar{c})}= 0.05\sqrt{a_0^D}$. 

We concider here these two values of $a_1$ taking the $\delta_{(c,\bar{c})} $ 
as more or less free parameter.

The two curves provided the fits of E791 and WA92 
pion beam experiment data \cite{e791,wa92}are presented in Fig.1 with two sets of parameters 
discussed above as well as the results of two other approaches \cite{lichoded,arakelyan}.

\begin{figure}[htb]
\begin{center}
\parbox[b]{8cm}{\psfig{scale=0.4,file=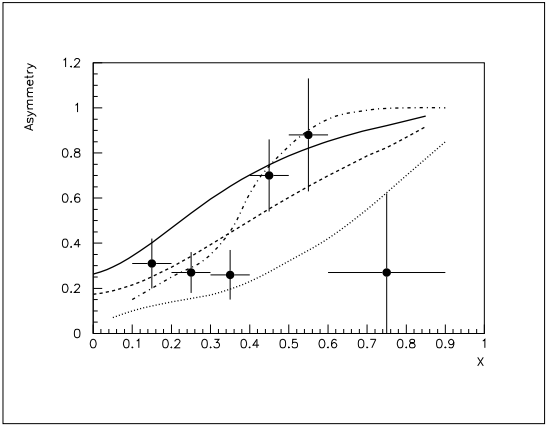}}\\
\caption{$D^-/D^+$ asymmetry measured in WA89 and theoretical calculations:
solid line corresponds to the following set of QGSM parameters $a_1=10,
\delta_(c,\bar{c})=0,01$; dashed line is a result of QGSM 
fit with  $a_1=4,\delta_{(c,\bar{c})}=0$; dashed-dotted line is the result
of \protect\cite{lichoded} and dotted line corresponds to A(x) predicted 
in \protect\cite{arakelyan}.}
\end{center}
\end{figure}

 It should be mentioned 
that the smaller fraction of charmed sea quarks was taken into 
account ($\delta_{(c,\bar{c})}=0.01\sqrt{a_0^D}$ for to 
describe both $D^-/D^+ and D_s^-/D^+_s$ asymmetries in 
$\Sigma^-A$ reaction instead of $\delta_{(c,\bar{c})}=0.05\sqrt{a_0^D}$
in the case of pion beam at the E791 experiment.


\subsection{The $\Lambda_c$/$\bar{\Lambda_c}$ Asymmetry.}

The asymmetry between the spectra of $\Lambda_c$ and $\bar{\Lambda_c}$ 
measured in $\Sigma^-A$ collisions can be easy obtained practically in the
same calculations described here. What is important, the leading 
$\Lambda_c$ 
baryon is formated from single d-quark of projectile particle. No diquark 
from $\Sigma^-$ hyperon takes part in leading charm baryon production. 
It allows us to take the results of our calculation for D-meson 
production at $p_L$= 600 GeV/c and to 
compare with the $\Lambda_c$/$\bar{\Lambda_c}$ asymmetry measured in 
in $\Sigma^-A$ collisions  in the  E781 experiment 
\cite{selex}.The parameter $a_1$ means in this case a parametrization 
parameter for the density
of uc-diquark in the string fragmentation, it can be different than $a_1$ 
taken for D-meson production. The energy of interaction has to be also 
changed. The parameter of the fraction of sea charm quarks , 
$\delta_{(c,\bar{c})}$, must be of the same value as for D-meson 
calculations because it doesn't know what leading particle is produced.

 The asymmetry between the spectra of $\Lambda_c$ and $\bar{\Lambda_c}$ measured in
$\Sigma^-A$ collisions at $p_L$= 600 GeV/c is shown in Fig.2.

\begin{figure}[htb]
\begin{center}
\parbox[b]{8cm}{\psfig{scale=0.4,file=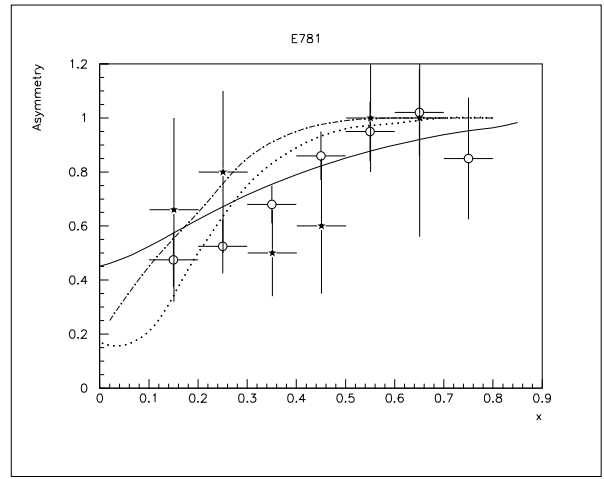}}\\
\caption{Asymmetry between $\Lambda_c$ and $\bar{\Lambda_c}$ spectra 
obtained in the E781 experiment 
(empty circles) \protect\cite{selex} and in the WA89 experiment (black stars) 
\protect\cite{wa89}; the preliminary QGSM curve (solid line) 
corresponds to the following set of 
parameters $a_1=25,\delta_{(c,\bar{c})}=0.01$; dashed-dotted line is 
the result of \protect\cite{lichoded} and dotted line corresponds 
to A(x) predicted in \protect\cite{arakelyan2}.} 
\end{center}
\end{figure}

The complete calculations carried out with the fragmentation 
function written for $\Lambda_c$ and $\bar{\Lambda_c}$ production 
with proton beam
give the good description of data with the value of diquark fragmentation parameter 
$a_f^{\Lambda_c}=0.008$ (see Fig.3).

\begin{figure}[htb]
\begin{center}
\parbox[b]{8cm}{\psfig{scale=0.4,file=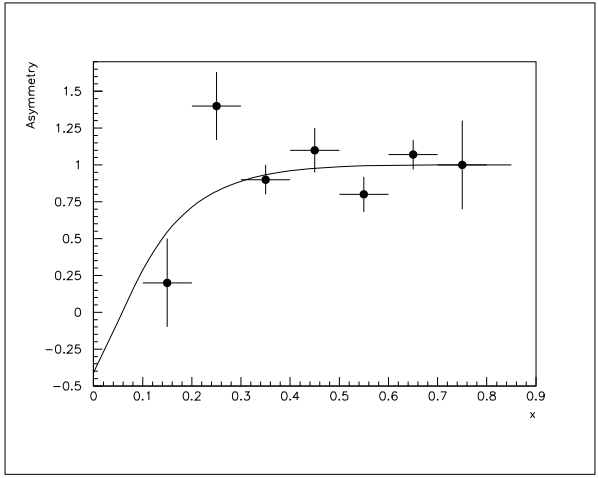}}\\
\caption{Asymmetry between $\Lambda_c$ and $\bar{\Lambda_c}$ spectra 
obtained in the E781 experiment with the proton beam 
(black circles) \protect\cite{selex}; the complete QGSM calculation (solid line).} 
\end{center}
\end{figure}

\section{CONCLUSIONS}

We can conclude here only that the data of the WA89 experiment on leading 
asymmetry in charmed meson production at $\Sigma^-A$ interactions
 can be described within the framework of Quark-Gluon String Model 
with the same  asymmetry parameter $a_1=10$ as  E791 data 
for $\pi^-A$ reaction. Good statistics on $D^-/D^+$ and $D_s^-/D^+_s$
production is needed to estimate the comparable influence of light,strange
and charmed sea quark fractions for different beams. 
$D^-/D^+$ and $D_s^-/D^+_s$ asymmetries measured with  $\Sigma^-$ 
beam are discribed with the different weight of charmed quark sea of interacting 
hyperon ($\delta_{(c,\bar{c})}$=0.01) than 
it was done for $\pi^-$ beam interaction ($\delta_{(c,\bar{c})}$=0.05).
$D_s^-/D_s^+$ asymmetry is higher than $D^-/D^+$ asymmetry because 
strange quark pairs suppressing the asymmetry at $D_s$ production have 
lower weight in quark sea of hyperon than ordinary $d\bar{d}$ pairs 
which cause the suppression of $D^-/D^+$ meson asymmetry.
Data of the E781 experiment on charmed baryon production 
asymmetry measured in $\Sigma^-A$ collisions can be also
described within the framework of Quark-Gluon String Model with 
the asymmetry parameter $a_1=25$ for uc-diquark density in d-quark 
string fragmentation. The asymmetry between the spectra of $\Lambda_c$ 
and $\bar{\Lambda_c}$ doesn't grow so rapidly with $x_F$ 
as it was predicted in other approaches. The rapid growing of asymmetry
in the case of proton beam can be explained by the important role of
ud-diquark fragmentation into $\Lambda_c$.


\begin{thebibliography}{99}
\bibitem{wa89}
WA89 Collaboration, M.I.Adamovich {\it et.al.},
Europian Phys. J.{\bf C8}(1999)593.
\bibitem{selex}SELEX Collaboration,M.Iori  {\it et.al.} these proceedings, 
hep-ex/0009049(2000).
\bibitem{e791}
E791 Collaboration, Aitala {\it et.al.}, Phys.Lett.{\bf B411}(1997)230.
\bibitem{wa92}
WA92 Collaboration, M.I.Adamovich {\it et.al.},
Nucl.Phys.{\bf B495}(1997)3.
\bibitem{qgsm}
Kaidalov A.B., Phys.Lett.{\bf 116B}(1982)459.
\bibitem{pionbeam}Piskounova O.I., Nucl.Phys.(Proc.Suppl.){\bf B50}(1996)179,
Phys.of At.Nucl.  {\bf 60}(1997)439.
\bibitem{prev papers}Piskounova O.I.,preprint FIAN-140,1987. 
\bibitem{kaidalov}Kaidalov A.B.,Sov.J.Nucl.Phys. {\bf 45}(1987)1450.
\bibitem{kaons}Kaidalov A.B. and Piskounova O.I., 
Phys.of At.Nucl. {\bf 41}(1985)1278.
\bibitem{charm}Kaidalov A.B. and Piskounova O.I., 
 Sov.J.Nucl.Phys. {\bf 43}(1986)994; Piskounova O.I.,Phys.of At.Nucl. 
{\bf 56}(1993)1094.
\bibitem{lichoded} 
Likhoded A.K. and Slabospitsky S.R.,Phys.of At. Nucl.{\bf 60}(1997)981.
\bibitem{arakelyan}Arakelyan G.H., Phys.Atom.Nucl.61(1998)1570.
\bibitem{arakelyan2}Arakelyan G.H., hep-ph/9906544 (1999).
\end{thebibliography}
\end{document}